\documentclass[preprint,10pt]{aastex}

\shorttitle{Ionization Mechanism in NGC~4569}
\shortauthors{Gabel \& Bruhweiler}

\begin{document}

\title{The Central Starburst and Ionization Mechanism in the LINER/H II Region Transition Nucleus in NGC~4569}
\author{J. R. Gabel and F. C. Bruhweiler}
\affil{Physics Department, The Catholic University of America,
    Washington, DC 20064}
\email{gabel@iacs.gsfc.nasa.gov}

\begin{abstract}
     We present a comprehensive study to determine if the LINER/\ion{H}{2} region transition spectrum in NGC~4569 can be generated solely by photoionization by the nuclear starburst.  A review of the multiwavelength data from the literature reveals no additional sources that contribute to the ionization.  We find that the young starburst dominating the UV emission is distinct from the nuclear population of A supergiants identified in the optical spectrum by Keel (1996).
Spectral synthesis analysis provides constraints on the physical nature of the starburst, revealing a 5-6~Myr, approximately instantaneous starburst with subsolar metallicity.  These results are used to model the spectral energy distribution of the ionizing continuum.  Luminosity constraints place limits on the steepness of the extinction curve for the young starburst.  The Savage \& Mathis (1979) curve satisfies all luminosity constraints and the derived reddening is similar to the emission line reddening.  These results imply extreme conditions in the nuclear starburst, with $\sim$5x10$^4$ O and B stars compacted in the inner 9\arcsec~x~13\arcsec region of the nucleus.
Using photoionization analysis and employing all observational constraints on the emission line gas, we find very specific conditions are required if the spectrum is generated solely by stellar photoionization.  At least two spatially distinct components are required - a compact region with strong \ion{O}{3} emission and an extended, low density component emitting most of the \ion{S}{2} flux.  A high density component is also needed to generate the \ion{O}{1} flux.  Additionally, a limited contribution from Wolf-Rayet stars to the ionizing SED is necessary, consistent with the results of Barth \& Shields (2000).  We present a physical interpretation for the multi-component emission line gas.
\end{abstract}


\section{Introduction}

     One of the most basic properties of low ionization emission line galactic nuclei, their ionization and excitation mechanism, is not currently well understood.  These objects include low ionization nuclear emission line regions (LINERs), which have optical spectra dominated by forbidden lines from low ionization species ([\ion{O}{1}]~$\lambda$6300, [\ion{S}{2}]~$\lambda$6726, [\ion{N}{2}]~$\lambda$6584), and ``transition objects", with optical line ratios intermediate to \ion{H}{2} regions (dominated by Balmer lines and sometimes [\ion{O}{3}]~$\lambda$5007) and LINERs.  Various ionization mechanisms have been proposed for these nuclei, including photoionization by a low luminosity active galactic nucleus (AGN) \citep{ferl83}, stellar photoionization \citep{terl85,fili92,bart00}, and shock-heating \citep{dopi95}, however, the true nature of only a few individual nuclei has been demonstrated conclusively \citep{bart96,hofi96,gabe00}.

     The physical nature of transition objects remains particularly uncertain.  While many LINERs have broad emission line components \citep{hofi97b} and compact, flat radio cores \citep{naga00} indicating they are {\it bona fide} AGNs, transition objects do not generally exhibit these features.  Thus, there is no direct evidence linking them to AGNs.  If stellar photoionization is the dominant mechanism in transition objects, an explanation for the uncharacteristically strong low ionization lines is necessary.  Recent studies have determined the essential features required for stars to generate such a spectrum: (a) The spectral energy distribution of the stars must be ``hard" compared to typical \ion{H}{2} regions, {\it i.e.}, it must have stronger relative emission beyond the \ion{He}{1} and \ion{He}{2} ionization limits.  A harder SED produces an extended partially ionized zone in the emission line region, thereby enhancing the LINER features \citep{pequ84,shie92,bart00}.  (b) Also, a low ionization parameter is required \citep{shie92,bart00}, which is a function of the starburst's luminosity, nebular density, and the geometry of the nucleus. 

     The transition object nucleus in NGC 4569 is an ideal object for testing a possible link between starburst activity and low ionization nebular emission.  Of all LINERs and transition nuclei with available UV data, NGC~4569 exhibits by far the most pronounced nuclear starburst activity. \citet{maoz98} found that the ionizing luminosity of the nuclear starburst in NGC~4569 is sufficient to generate the observed emission line luminosity.  \citet{bart00} demonstrated that a large population of very hot Wolf-Rayet stars could harden the integrated SED of a starburst enough to produce the partially ionized zone required to produce a transition object spectrum.  They also found that the estimated age of the stellar population in NGC~4569 roughly coincides with the W-R phase of evolution in a starburst.

     However, there are no observational signatures of W-R stars in the spectrum of NGC~4569 \citep{bart00}.  Additionally, W-R galaxies typically have emission line spectra characteristic of H II regions, rather than transition objects or LINERs \citep{guse00}.  Furthermore, it is unclear if the ionization parameter in NGC~4569 is sufficiently low to generate the observed spectrum.  Hence, the questions remain: Is the observed line emission in NGC~4569 generated solely by photoionization by the nuclear starburst?  If so, what physical conditions in the nebular gas and starburst are required?

     We have undertaken a detailed analysis of the emission in NGC~4569 to answer these questions, incorporating all available observational constraints.  In \S~2, we summarize the multiwavelength data of NGC~4569 from the literature and present the archival data used in our analysis.  In \S~3, we apply population synthesis analysis to these data to place constraints on the nature of the nuclear starburst and the emitted flux distribution.  Using the derived constraints on the SED and luminosity of the starburst, along with observational constraints on the emission line gas, we use photoionization modeling in \S~4 to determine if stellar photoionization can generate the observed transition object spectrum emitted by NGC~4569.  Finally, we discuss the implications of our results in \S~5.  

\section{Nuclear Data and Multiwavelength SED in NGC~4569}

     In the following, we describe the optical and UV data used in our analysis and summarize results obtained from the literature for several bandpasses.  These multiwavelength observations are crucial for determining the physical mechanisms that power the nuclear emission.

     For our study, we adopt a distance of 16.8~Mpc to NGC~4569, given by \citet{deva91}.  Due to the anomalous radial velocity of NGC~4569, $v=-$223~km~s$^{-1}$, there has been much debate concerning this distance, and it remains highly uncertain \citep[and references therein]{stau86}.  We explore the affects of this uncertainty on our results in \S~5.

\subsection{Spectrum and Morphology of the Emission Line Region: A LINER/H II Region Transition Object}

     NGC~4569 was observed as part of the Palomar survey of nearby galactic nuclei by \citet{hofi97a} using the Hale~5m telescope with a 2$\arcsec$~x~4$\arcsec$ aperture ($\approx$160~x~320~pc).  Based on the observed emission line ratios, [O I]~$\lambda$6300/H$\alpha$, [\ion{S}{2}]~$\lambda$6726/H$\alpha$, [\ion{N}{2}]~$\lambda$6584/H$\alpha$, and [\ion{O}{3}]~$\lambda$5007/H$\beta$, they classified NGC~4569 as a LINER/\ion{H}{2} region transition object nucleus.  The observed line luminosities and estimated uncertainties measured by \citeauthor{hofi97a} are given in Table~1.  
Using the \citet{sava79} extinction curve and assuming an H$\alpha$/H$\beta$ flux ratio of 2.9, which is typical of emission line nuclei \citep{oste89}, we derive a reddening of $E(\bv)=$0.50 ($\pm$0.15) for the emission lines.  Dereddened line luminosities are presented in column~3 of Table~1.  The quoted uncertainties were derived from the uncertainties in the flux measurements and the reddening, using propagation of errors.  The uncertainty in the extinction curve in NGC~4569 introduces an additional uncertainty and is examined in more detail in \S~5.2.

     The morphology of the nuclear line emission in NGC~4569 is delineated in the {\it Hubble Space Telescope (HST)} WFPC2 narrowband imagery presented by \citet{pogg00}.  The continuum subtracted H$\alpha$ (F656N) and [\ion{O}{3}]~$\lambda$5007 (F502N) images shown in Figures 1c and 3b of their study reveal a bright nuclear core with total extent $<$~1$\arcsec$ (80~pc).  There is also a more extended, diffuse nebular emission component apparent in both images.  Examination of the spectral bandpass of the F656N filter reveals little contamination from the [\ion{N}{2}]~$\lambda\lambda$6548,6584 lines at the blueshift of NGC~4569, and is therefore a good measure of the H$\alpha$ emission.  For each image, \citeauthor{pogg00} have measured the flux emitted in the inner $R=$0$\arcsec$.23 (20 pc) core of the nucleus, as well as the more extended nuclear region, $R=$2$\arcsec$.5 (250 pc).  Their results give a measure of the compactness of the line emission:  $L_{20pc}$/$L_{250pc}=$0.31 for H$\alpha$ and $L_{20pc}$/$L_{250pc}=$0.74 for [\ion{O}{3}].  These flux ratios provide valuable constraints for our photoionization analysis in \S~4.
     
\subsection{UV Data: Evidence for a Compact Nuclear Starburst in NGC~4569}

     We obtained Faint Object Spectrograph (FOS)/{\it HST} UV data from the Space Telescope Science Institute (STScI) data archives.  These data were originally published by \citet{maoz98}.  Two exposures with the G130H grating, and one each with the G190H and G270H gratings were used. These data have a spectral resolution of $\Delta\lambda\sim$1-2~\AA\ .  Details of the observations are summarized in Table~2. The two G130H grating exposures were coadded by weighting them according to their exposure times.  A combined spectrum was then created from the three spectra and corrected for the systemic velocity of the galaxy.
 
     All exposures were taken through the 0$\arcsec$.86 square aperture of the FOS.  In a study of {\it HST}/WFPC2 images taken with the F218W filter ($\lambda_{central}=$2200~\AA  ), \citet{bart98} found the UV nucleus of NGC~4569 to be very compact though clearly resolved, with a FWHM of 0$\arcsec$.16~x~0$\arcsec$.11 (13~x~9~pc$^2$).  Thus, the FOS data sample the entire compact UV emitting region.  Furthermore, we find that the UV flux of the compact nucleus measured from the {\it HST} imagery by \citeauthor{bart98} matches the {\it International Ultraviolet Explorer (IUE)} spectrum obtained through the 10$\arcsec$~x~20$\arcsec$ aperture \citep{keel96} within the noise of the data.  Hence, our comparison shows no significant UV emitting sources, other than the compact nuclear starburst, are contributing to the ionization of the emission line gas.

     \citet{maoz98} found that the FOS UV spectrum of NGC 4569 is dominated by a very young starburst.  The resonance doublets of \ion{C}{4}~$\lambda\lambda$1548,1551, \ion{N}{5}~$\lambda\lambda$1239,1243, and \ion{Si}{4}~$\lambda\lambda$1393,1403 exhibit P~Cygni profiles, indicative of the strong stellar winds produced in O and B stars.  \citeauthor{maoz98} found that the line profiles of the wind features in NGC~4569 are similar to the starburst in NGC~1741, which has a derived age of $\approx$5~Myr \citep{cont96}.  They also identified several photospheric features that appear in young, hot stars.  

      Numerous interstellar lines, both Galactic and intrinsic to the host galaxy, are also apparent in the UV spectrum of NGC~4569 \citep{maoz98}.  Strong features from the multiplets of \ion{C}{1} near $\lambda$1277, $\lambda$1280, $\lambda$1329, and $\lambda$1561 were detected at the systemic velocity of NGC~4569, which implies the starburst spectrum is seen through a large column of cool, possibly high density gas intrinsic to NGC~4569.  Thus, large extinction of the UV spectrum due to dust is expected.  Further evidence for a high extinction of the starburst comes from the flatness of the observed UV spectrum between 1200-3000~\AA\ (\citeauthor{maoz98}), since an unreddened population of O and B stars would have a sharply rising continuum toward shorter wavelengths.  

     Besides the red-shifted emission lobes associated with the stellar P~Cygni resonance profiles, there are no other apparent emission lines in the UV FOS spectrum of NGC~4569 in the 1200-3000~\AA\ range.  Notably, there are no detectable stellar emission features that might indicate the presence of Wolf-Rayet stars, such as \ion{He}{2}~$\lambda$1640 or \ion{C}{3}]~$\lambda$1909, nor are there UV nebular emission lines.  Upper limits on nebular emission line fluxes were determined by \citet{maoz98} and are included with our optical fluxes in Table~1.

\subsection{The Optical Counterpart to the UV Starburst}

     The continuum flux of the optical emission coinciding with the UV starburst can be derived from analysis of {\it HST}/WFPC2 continuum images.  Combined with the UV spectrum, this provides constraints in deriving the extinction of the central starburst (see Section 3.3).  Exposures taken with the F547M ($\lambda_{central}=$5470~\AA) and F791W ($\lambda_{central}=$7920~\AA) filters were used because their bandpasses are sufficiently narrow to exclude contribution from strong nebular line emission, thereby providing fairly reliable stellar continuum fluxes.  Although the total CCD detector counts are high, data for the bright core in each of these images remain unsaturated.  We integrated the total fluxes in these images over a 0$\arcsec$.35~x~0$\arcsec$.35 spatial region corresponding to the approximate size of the UV nucleus as determined from the WFPC2 F218W image \citep{bart98}.  In Figure~1, we show the F547M image and delineate the region on the sky used for our core flux measurement.  Specific fluxes were derived using the conversions provided by Holtzman et~al. (1995), assuming a flat flux distribution.  This yields $F_{5470\AA}=$5.0x10$^{-15}$~ergs~s$^{-1}$~cm$^{-2}$~$\AA^{-1}$ and $F_{7920\AA}=$2.6x10$^{-15}$~ergs~s$^{-1}$~cm$^{-2}$~$\AA^{-1}$.

     In a ground-based study of NGC~4569 using a 2$\arcsec$~x~4$\arcsec$ aperture with the 2.5m Isaac Newton Telescope, \citet{keel96} deduced that the optical stellar continuum is dominated by a population of A-type supergiants from the strong narrow Balmer absorption features.  These stars point to a much older population ($>$~15~Myr) than implied by the UV data.  Figure~1 reveals that the optical stellar continuum emission is clearly more extended than the compact UV starburst.  We have measured the flux in the inner 2$\arcsec$~x~4$\arcsec$ region of the F547M image and find that $\leq$20\% of the optical flux coincides with the UV source, {\it i.e.}, the inner 0$\arcsec$.35~x~0$\arcsec$.35 region measured above.  Therefore, the population of A supergiants must be in the extended component seen in Figure~1 and cannot be directly associated with the compact starburst.  This indicates that the nucleus of NGC~4569 contains at least two distinct young stellar populations: (1) the very young, very compact UV core and (2) a spatially more extended population dominated by A supergiants, with few, if any, young O and B stars present.

\subsection{Multiwavelength Observations}

      Although the UV spectral data and imagery imply that a very young stellar population is dominating the emission in NGC~4569, this does not rule out the possibility that an obscured AGN is present, which can contribute to the ionization in the emission line gas.  Below, we summarize multiwavelength observations obtained from the literature to test this possibility.

{\it Radio Emission}

     There are no signatures of an AGN in radio observations of NGC~4569.  High angular resolution radio data obtained with the Very Large Array (VLA) at 6~cm and 20~cm reveal only weak, extended ($\approx$4$\arcsec$, 320~pc) emission in the nuclear region, with integrated fluxes of $F_{6cm}=$1.5~mJy and $F_{20cm}=$10~mJy \citep{neff92,humm87}.  No compact core component is detected in this data, with an upper limit on any unresolved flux of $F_{6cm}<$0.4~mJy at 1$\arcsec$ angular resolution \citep{neff92}.  The spectral index of $\alpha_{20-6}=-$1.5, derived by extrapolating between the fluxes at $\lambda=$20 and 6 cm and assuming $F_{\nu}\propto\nu^{-\alpha}$, is much steeper than the slope expected for pure free-free emission in an \ion{H}{2} region, $\alpha_{ff}$, revealing a substantial non-thermal contribution.  We estimated the contribution of free-free emission to the observed flux using the H$\alpha$ emission line luminosity given in Table~1, following the method of \citet{oste89}.  Assuming all of the H$\alpha$ emission is generated by photoionization in a nebula at $T_e =$10$^4$~K gives $\sim$0.7~mJy for the free-free flux, or roughly one-half of the observed value at 6~cm.  Subtracting this from the observed fluxes implies a spectral index of $\alpha_{20-6}=-$1.9 for the non-thermal emitting component.

     The steep slope of the radio emission in NGC~4569 may be the signature of supernovae remnants (SNRs).  For example, a survey of SNRs in M31 shows a range of spectral slopes, $\alpha_{20-6}\sim-$0.35 to  $-$1.6 \citep{dick84}.  Assuming the non-thermal component of radio emission in NGC~4569 is from SNRs and using an SNR luminosity at 6~cm of $L_{6cm}=$1.1x10$^{24}$~ergs~s$^{-1}$~Hz$^{-1}$ (\citeauthor{dick84}), approximately 100-200 SNRs are required in the nucleus of NGC~4569, depending on the contribution of the free-free thermal emission.

{\it Far-Infrared Emission}

     Far-infrared (FIR) fluxes at $\lambda=$~12, 25, 60, and 100~$\mu$m are available from {\it IRAS} observations \citep{neff92}.  These data, at $\sim$30$\arcsec$ resolution, sample an extended region in NGC~4569.  Since there are no signatures of a central AGN in the radio data, the FIR emission most likely arises from warm dust.  Assuming most of the nuclear emission is reprocessed as thermal dust emission, the total FIR luminosity, $L_{FIR}=$2x10$^{43}$~ergs~s$^{-1}$, provides an upper limit on the bolometric luminosity of the starburst.  We have compared the FIR spectral energy distribution (SED) in NGC~4569 to blackbody flux distributions at various temperatures and find that at least two dust temperatures are required to fit the data.  This is not surprising considering the large spatial region sampled by the {\it IRAS} data.  The warmer dust may be associated with the inner nuclear region, close to the energetic starburst and possibly in the same gas giving rise to the observed \ion{C}{1} absorption, while the cooler dust may be located in more extended regions.

{\it X-Ray Emission}

     Previous studies of the nuclear X-ray emission in NGC~4569 also reveal no signatures of an AGN.  \citet{tera00} found no compact hard X-ray emission in {\it ASCA} observations, placing an upper limit of $L_{2-10~keV}<$3x10$^{39}$~ergs~s$^{-1}$ on any unresolved core luminosity.  Thus, if an AGN capable of powering the nebular line emission in NGC~4569 is present, it would have to be heavily obscured.  A compact component has been detected at soft X-ray energies with {\it ROSAT}/HRI, $L_{0.1-2.4~keV}=$6x10$^{39}$~ergs~s$^{-1}$ \citep{colb99}, and with {\it ASCA}, $L_{0.5-2.0~keV}=$6x10$^{39}$~ergs~s$^{-1}$ (\citeauthor{tera00}).  The $\sim$4$\arcsec$ resolution of {\it ROSAT}/HRI imposes the most rigid constraint on the size of any compact emitting region, but still samples the entire spatial region of the compact stellar core, the nebular gas, and the radio emitting source.  Spectral analysis of the soft {\it ASCA} data by \citeauthor{tera00} indicates that it is not consistent with scattering of non-thermal emission by an obscured AGN.  Combining this result with the low upper limit on the compact hard X-ray flux, they concluded that the unresolved soft X-ray emission is associated with the nuclear starburst rather than an AGN.

     Young starbursts are known to emit soft X-rays.  The combined emission from SNRs, interstellar gas heated by supernovae and stellar winds, and X-ray binaries can all contribute to the integrated X-ray emission in a starburst \citep[and references therein]{fabb89}.  Unfortunately, distinguishing between these mechanisms in NGC~4569 by spectral modeling is not possible with the available data.  The core luminosity in NGC~4569 is consistent with other starburst nuclei, which are typically in the 10$^{39}$~ergs~s$^{-1}$ range (\citeauthor{fabb89}).  However, the stellar population and the X-ray source are more compact in NGC~4569 than in typical starburst nuclei.  
 
{\it The Broadband Spectral Energy Distribution}

     In Figure~2, the observed SED of the nuclear emission in NGC~4569 from radio to soft X-ray energies is plotted.  UV fluxes were obtained from the FOS spectrum.  The optical data are for the inner 0$\arcsec$.35~x~0$\arcsec$.35 core of emission derived in \S~2.3, while the radio data points are upper limits to any unresolved flux.  Due to the low angular resolution of {\it IRAS}, the FIR data represent upper limits to the nuclear core flux.  The soft X-ray SED in Figure~2 was derived by fitting a power law flux distribution ($F_{\nu}\propto\nu^{-\alpha}=-$1.5) to the observed compact (within $\approx$4$\arcsec$) {\it ROSAT}/HRI flux quoted in \citet{colb99}.  This was done only as an approximation and has no physical basis. 
     
     Qualitatively, the entire broadband nuclear SED is consistent with emission associated with a young stellar population.  Figure~2 shows that the emitted power is strongest in the optical-UV region, and possibly in the FIR (depending on the contribution from the inner nucleus), with relatively low luminosities at radio and X-ray frequencies.  The spectral signatures in the UV indicate a nuclear starburst is dominating the emission and the FIR luminosity can be understood solely as reprocessed stellar radiation with no indications that an AGN is present.   

\section{Spectral Population Synthesis of the Nuclear Starburst}

     To test if photoionization by the nuclear starburst can generate the LINER/\ion{H}{2} region transition spectrum emitted in NGC~4569, we must first establish the integrated spectral energy distribution and luminosity of the stars.  Thus, we employ spectral population synthesis analysis in this section to derive constraints on the evolutionary stage and physical conditions in the starburst.  These results are then combined with stellar atmosphere models to generate a model SED for subsequent photoionization analysis.  The derived flux distribution, along with observational constraints described in the previous section, place limits on the reddening, and hence the luminosity of the nuclear starburst.

\subsection{Description of the Modeling Code}

     In the following, we employ a population synthesis modeling code, CLUST\_FLUX, which generates UV spectra for a starburst of specified age, metallicity, and initial mass function (IMF).  CLUST\_FLUX computes the present day stellar population using the stellar evolutionary models of \citet{schl92}, \citet{sche93a,sche93b}, \citet{char93}, and \citet{meyn94}. These models predict the isochrones in luminosity - effective temperature parameter space ($L-T_{eff}$) for a given age and metallicity.  Combining these results with the IMF, the relative number of stars for all locations in the $L-T_{eff}$ plane is then fully specified.  Stellar evolutionary tracks that include the effects of mass loss are available for various metallicities, ranging from 0.05-2~Z$_\sun$, where Z$_\sun$ represents solar metallicity. 

     CLUST\_FLUX generates a composite model starburst spectrum using short wavelength {\it IUE} data (1170-2000~\AA ).  A weighted sum of individual stellar spectra is created from a library of {\it IUE} data compiled for the Milky Way and Large and Small Magellanic Clouds, which are classified according to effective temperature ($T_{eff}$) and surface gravity (log($g$)).  Thus, this library is a 3-dimensional grid of spectral data in $T_{eff}$, log($g$), and $Z$, covering three metallicities: Z$_\sun$, 0.25~Z$_\sun$ (LMC), and 0.1~Z$_\sun$ (SMC).  For stellar types not corresponding to exact matches in $T_{eff}-log(g)-Z$ in the {\it IUE} library, interpolated spectra are generated.  To correct for extinction, with often unknown, highly variable reddening laws for stars in the {\it IUE} data, all spectra have been matched to the flux distributions of Kurucz (1992) model atmospheres for corresponding $T_{eff}-log(g)-Z$.
   
     The UV spectral classification criteria of \citet{smit97,smit99}, aided with the UV line identifications of \citet{dean85} and \citet{bruh81}, were used to classify the UV stellar spectra for OB stars in the SMC and LMC.  In practically all cases, these classifications were in close accord with corresponding visual spectral types in all three galaxies.  High resolution {\it IUE} spectra of Galactic stars are available for the solar metallicity grid and have been binned into $\Delta\lambda=$0.25~\AA\ intervals.  Since there are few high dispersion {\it IUE} spectra for stars in the Magellanic Clouds, the stellar library of sub-solar metallicity SMC and LMC stars is comprised of low resolution ($\Delta\lambda$$\sim$7~\AA\ ) {\it IUE} spectra.
  
     Due to the large uncertainties involved in predicting the population of Wolf-Rayet stars in a starburst \citep{schm99}, we do not include them in the integrated model starburst spectra.  Thus, CLUST\_FLUX does not predict the strengths of the broad UV emission features that are signatures of Wolf-Rayet stars, such as \ion{He}{2}~$\lambda$1640, \ion{C}{4}~$\lambda$1550, and \ion{C}{3}]~$\lambda$1909.  However, emission from W-R stars can significantly alter the integrated ionizing emission in a starburst and has important consequences for photoionization analysis, as demonstrated by \citet{bart00}.  Therefore, we explore observational constraints on the W-R population in NGC~4569 separately in \S~3.2.

\subsection{Stellar Population Synthesis Results}

     Our method is to examine the effect of age on an instantaneous burst of star formation using standard model parameters and to determine the best fit to the spectrum of NGC~4569.  The effects of varying metallicity, the IMF, and star formation rate can then be explored to better constrain the physical nature of the starburst.  In the following analysis, we adopt the high mass-loss rates of the \citet{meyn94} evolutionary models.

{\it Age Constraints}

     We have generated model spectra for an instantaneous burst of star formation for a range of ages using the standard IMF of \citet{mill79} and solar metallicity.  Their IMF, $\xi_{MS}$, is defined by a three component power law function of mass, with spectral indices $\alpha_1=-$0.4, $\alpha_2=-$1.5, and $\alpha_3=-$2.3 in the mass ranges $M_1=$0.1-1~M$_\sun$, $M_2=$1-10~M$_\sun$, and $M_3=$10-120~M$_\sun$, respectively.

     The synthesized, normalized spectra in four far-UV spectral regions for instantaneous burst models with ages 0, 2, 4, 5, 6, and 8 Myr are plotted in Figure~3.  These show the spectra for the \ion{C}{4}, \ion{N}{5}, \ion{Si}{4}, and \ion{He}{2} features, as well as important photospheric diagnostic lines.  The normalized spectrum of NGC~4569 is superposed on all model spectra.  Due to the limited S/N in the FOS spectrum of NGC~4569, we have convolved the data with a $\sigma=$1.0~\AA\ Gaussian filter for clarity.  To reproduce the spectral resolution of the smoothed FOS data, all model spectra were smoothed using a Gaussian filter with $\sigma=$1.5~\AA , which reflects the best match between the spectral widths of the photospheric features in the model and observed spectra.  The uncertainties in our diagnostic line fits are dominated by uncertainties introduced by the photon noise in the spectrum of NGC~4569.

     In Figures 3a-c, the stellar wind profiles of the \ion{N}{5}, \ion{C}{4}, and \ion{Si}{4} resonance doublets are seen to vary dramatically with age between 0 and 8~Myr.  The observed \ion{N}{5}~$\lambda\lambda$1239,1243 feature in the spectrum of NGC~4569 does not exhibit the broad blue absorption wing seen in the very young starburst models, indicating an age of at least 4 Myr.  However, the red-shifted \ion{N}{5} emission lobe of the P~Cygni profile is clearly present in NGC~4569 and is best reproduced by the 5-6~Myr instantaneous burst models.  Since the \ion{Si}{4}~$\lambda\lambda$1394,1403 doublet feature is most pronounced in high luminosity supergiants, it is a very useful age discriminate.  Strong \ion{Si}{4} appears for only a short range of ages and does not occur in the youngest starbursts, as illustrated in Figure 3b.  The best-fit to the \ion{Si}{4} wind feature is 4-5~Myr, with a rigid upper limit of 6~Myr.  For later ages, the \ion{Si}{4} feature vanishes entirely.  The \ion{C}{4}~$\lambda\lambda$1548,1551 P~Cygni feature is strongest in the very youngest starbursts and weakens systematically with age (Figure 3c).  It is best matched by the 5 and 6~Myr models.  The subordinate line, \ion{N}{4}~$\lambda$1718, which exhibits a P~Cygni profile in spectral types O7 and earlier, is not present in NGC~4569.  Our models predict this feature will become undetectable in starbursts older than 4 Myr.  In Figure~3d, the spectrum for \ion{He}{2}~$\lambda$1640 is plotted.  Since the CLUST\_FLUX models do not include a contribution from W-R stars, the model predictions for this spectral region are similar for different ages.  No evidence for a broad \ion{He}{2}~$\lambda$1640 emission feature is evident in the spectrum of NGC~4569, thereby limiting the number of W-R stars in the starburst (see discussion below). 

     The strengths of the photospheric lines also provide important age diagnostics.  The \ion{Si}{3}~$\lambda\lambda$1295-1303 lines from the $^3$P-$^3$P$^0$ multiplet peak in strength in late B-type stars and, thus, increase through 8~Myr.  The \ion{Si}{3} lines that are uncontaminated by interstellar O~I~$\lambda$1302 and Si~II~$\lambda$1304 absorption are plotted in Figure~3a.  In the NGC~4569 spectrum, the strength and width of the \ion{Si}{3}~$\lambda$1294.5 multiplet line is inconsistent with the \ion{Si}{3}~$\lambda$1296.7 and \ion{Si}{3}~$\lambda$1298.9 lines and is therefore an unreliable diagnostic.  However, the \ion{Si}{3}~$\lambda$1296.7 and \ion{Si}{3}~$\lambda$1298.9 line strengths are matched by models with ages $\geq$~5~Myr.  The \ion{Si}{3}~$\lambda$1417 line, shown in Figure~3b, exhibits the same trend with age as the \ion{Si}{3}~$\lambda\lambda$1295-1304 multiplet and indicates an age of 5~Myr or older in NGC~4569.  In Figures 3a and b, the \ion{C}{3}~$\lambda$1247 and \ion{C}{3}~$\lambda\lambda$1426-1428 line strengths also vary with age.  The strength of each of these lines in NGC~4569 implies an age $\geq$~4~Myr.  At later ages, the lines show little variation and provide no upper limit for the ages of interest here.  In Figure~4, we plot the model and observed spectra for 1890-1990~\AA\.  This samples numerous \ion{Fe}{3} lines (marked on the top of Figure~4) that become prominent in B~supergiants, thus providing an additional age diagnostic.  The \ion{Fe}{3} multiplets are seen to increase in strength with later ages. The spectrum of NGC~4569 is best matched at these wavelengths by ages of 5~Myr and older.

     The combined fits to the stellar wind and photospheric features indicate an age of 5-6~Myr for the starburst.  Particularly strong limits come from the presence of the \ion{N}{5} wind feature, indicating the starburst is $\leq$~6~Myr, and the strength of the short-lived  \ion{Si}{4} wind line, implying an age of 4-6~Myr.  The photospheric line diagnostics imply an age of at least 5 or 6 Myr for NGC~4569.

{\it Initial Mass Function}

     Model spectra were also generated using different values for the IMF, star-formation rate, and metallicity in an attempt to constrain these parameters in NGC~4569.  To test the IMF, we produced models using both a flatter ($\alpha=-$1.5) and steeper ($\alpha=-$3) IMF relative to $\xi_{MS}$ for the stellar masses $M\geq$10~M$_\sun$.  Like the previous models, these spectra were generated for an instantaneous burst of star formation using solar metallicity.

     A flatter IMF results in a spectrum that is more heavily weighted with features from the most massive main-sequence (MS) stars.  Thus, features that arise primarily in the most evolved stars are enhanced, resulting in stronger P~Cygni features at all ages relative to the models using $\xi_{MS}$.  Conversely, we find the predicted strengths of the photospheric lines appearing in the spectrum of NGC~4569 are diminished for ages younger than 6 Myr in the flat IMF models.  Thus, if the IMF in NGC~4569 is flatter than $\xi_{MS}$, the observed strengths of the photospheric features indicate an earlier age than derived previously.

     Of course, the models having a steep IMF give opposite results.  This provides a constraint on the IMF in NGC~4569.  Specifically, since the observed strength of the \ion{Si}{4} wind feature in NGC~4569 approaches the maximum strength predicted by the standard models (Figure~3b), the IMF in NGC~4569 must be at least as flat as $\xi_{MS}$.  Otherwise, the \ion{Si}{4} feature would be underestimated in models for all ages.

{\it Star Formation History}

     The star formation history of the starburst in NGC~4569 is also constrained by the observed strength of the \ion{Si}{4} wind feature.  Since a pronounced \ion{Si}{4} wind feature occurs over only a short age interval and since the strength of this feature in NGC~4569 equals the maximum predicted strength for an instantaneous burst, the duration of star formation must be relatively short.  For star formation lasting longer than about 3 Myr, the \ion{Si}{4} wind feature would be diluted and weaker than observed in NGC~4569.  A similar argument applies in limiting the contribution from multiple bursts of star formation.  Thus, the starburst in NGC~4569 is consistent with a single burst of approximately instantaneous star formation.  \citet{bart00} reached a similar conclusion based on the small spatial extent of the starburst.
  
{\it Metallicity}

     Evolutionary models show that the mass-loss rates of massive stars are highly dependent on metallicity.  Since mass-loss rates determine the stellar evolutionary tracks in the HR diagram and the rate of progression along those tracks \citep{schl92}, the constituents of a stellar population are very sensitive to metallicity.

     With the LMC and SMC data in our stellar library, it is possible to test the effect of metallicity on a starburst spectrum.  We have generated spectra for both $Z=$Z$_\sun$/4 (LMC) and $Z=$Z$_\sun$/10 (SMC) for comparison with our standard models.  The most striking difference is the decrease in strength of the wind features with decreasing metallicity.  For example, the 5~Myr, Z$_\sun$/10 model spectrum significantly underestimates the strengths of all three wind features observed in NGC~4569.  Although the observed strengths of the \ion{N}{5} and \ion{C}{4} features can be reproduced by $Z=$Z$_\sun$/10 models with earlier ages, the \ion{Si}{4} line cannot.  Additionally, the weakness of the photospheric features in the low metallicity, young burst models are incompatible with the spectrum of NGC~4569.  This implies $Z>$Z$_\sun$/10 for NGC~4569.  Due to insufficient S/N in the spectrum of NGC~4569, we cannot place more rigid constraints on the metallicity.

{\it Wolf-Rayet Stars}

     The burst age of 5-6~Myr derived from our population synthesis analysis coincides with an age where large numbers of W-R stars may be present, although this depends strongly on the precise age and metallicity of the starburst \citep{meyn94,schm99,guse00}.  As discussed by \citet{bart00}, no W-R emission features are detectable in the optical spectrum obtained by \citet{hofi97a}.  Furthermore, we find no broad \ion{He}{2}~$\lambda$1640 ($W_{\lambda}<$1.0~\AA) or \ion{C}{3}]~$\lambda$1909 emission, the UV signatures of W-R stars, in the FOS spectrum of NGC~4569.  Additionally, the  \ion{C}{4}~$\lambda$1550 wind feature was well matched by our population synthesis models using only normal O and B stars (Figure~3c), which limits the contribution to this feature from W-R stars.
   
     Since the W-R population in a starburst is extremely sensitive to age and metallicity, the absence of detectable W-R features in NGC~4569 provides further constraints on the starburst.  We employ the population synthesis results of \citet{sche98}, who derive the strength of the \ion{He}{2}~$\lambda$1640 emission feature in an instantaneous starburst as a function of age and metallicity.  Their models employ the same stellar evolutionary tracks used in CLUST\_FLUX and incorporate a full treatment of the relative numbers of the different classes of W-R stars.  In figure~10 in \citeauthor{sche98}, they give the predicted equivalent width of the W-R \ion{He}{2}~$\lambda$1640 
emission line as a function of starburst age for various metallicities.  This figure predicts $W_{\lambda}$$\sim$6~\AA\ for a 5-6~Myr instantaneous starburst with solar metallicity,
which is six times greater than our measured upper limit.  The \ion{He}{2} emission is systematically reduced for lower metallicities, with $W_{\lambda}$$\sim$1~\AA\  in the 5~Myr $Z=$Z$_\sun$/2.5 model, and smaller for older starbursts.  Thus, if the spectral synthesis model predictions of the W-R population is correct, this indicates $Z\la$Z$_\sun$/2.5 in NGC~4569.  

{\it The SED, Luminosity, and Mass of the Nuclear Starburst}

     Our spectral synthesis analysis indicates that a 5-6~Myr instantaneous burst of star formation with Z$_\sun$/10$<Z\la$Z$_\sun$/2.5 and a \citet{mill79} IMF matches the overall spectrum of NGC~4569 quite well. CLUST\_FLUX is similar to the well established population synthesis code of \citet{leit99}, STARBURST99.  We have computed numerous comparison models and find the spectra generated by the two models compare favorably.  Specifically, they give identical results for the best fit to the starburst in NGC~4569.  These results can now be used to construct a model of the ionizing flux distribution for the starburst using STARBURST99.  Their models employ the model atmospheres of \citet{leje97}, based on the plane-parallel, LTE, line-blanketing models of \citet{kuru92}, for the OB stars and those of \citet{schm92} for the W-R stars.  

     In Figure~5, we plot the 5 and 6~Myr model SEDs from STARBURST99 computed for $Z=$Z$_\sun$/2.5, consistent with our spectral synthesis results.  A dramatic change in the shape of the SED occurs at wavelengths shortward of the He$^+$ ionization edge due to the disappearance of most of the hot W-R stars shortly after 5~Myrs.

      As discussed in \S~2, the strong interstellar \ion{C}{1} and the flat observed UV continuum in the FOS spectrum implies a substantial extinction of the nuclear starburst emission.  We adopted the extinction curve of \citet{sava79} to correct for this.  We address limits on the extinction law in NGC~4569 in more detail in \S~5.2.  Comparison to the shape of the model SED between 1200-1600~\AA\ implies a reddening of E($\bv$)=0.5 of the starburst emission.  This value is consistent with the reddening derived for the emission lines from the Balmer decrement.  Integrating the model flux distribution then gives L$_{Bol}=$1.5x10$^{43}$~ergs~s$^{-1}$ and L$_{\geq 1 Ryd}=$10$^{42}$~ergs~s$^{-1}$ for the bolometric and ionizing luminosities, respectively.  

     The nuclear starburst in NGC~4569 is quite remarkable.  Integrating the IMF of \citet{mill79} between 0.1~M$_{\sun}$$\leq$M$_\star \leq$~120~M$_{\sun}$ gives a total mass of 3x10$^7$~M$_\sun$ in the stellar population.  There are $\sim$5x10$^4$ luminous O \& B stars (assumed to be those with MS masses M$_\star \geq$15 M$_\sun$).
Thus, for the massive stars alone, the cluster has a stellar density of about 10~pc$^{-3}$.  We note that these results are sensitive to the uncertain extinction law of the starburst (see discussion in \S~5.2).  Comparison to the starburst nucleus in NGC~7714 \citep{gonz99} reveals NGC~4569 has about the same number of OB stars compacted in a volume that is 3000 times smaller.  Though extreme, the stellar densities in NGC~4569 are not unprecedented.  For example, the stellar density is believed to be similar in the Galactic center \citep[and references therein]{genz87}.  Furthermore, the densities and luminosities in the super star clusters in the Irregular galaxies NGC~1569 and NGC~1705 \citep{ocon94} are comparable. Interestingly, these latter objects are considered to be young globular clusters.  This may indicate that the starburst in NGC~4569 is a gravitationally bound system.

\section{Stellar Photoionization Modeling of the Transition Nucleus in NGC~4569}

     The primary question we are addressing is: Can ionization by a pure starburst explain the LINER/\ion{H}{2} region transition object spectrum in NGC~4569?  Now that we have a model of the SED and luminosity of the ionizing emission of the starburst, we can explore this using photoionization analysis.  In the following, we present models using our derived SED and all the available observational constraints on the physical conditions of the emission line gas.

\subsection{Modeling Method, Assumptions, and Parameters}

     Models were generated using the photoionization code CLOUDY \citep{ferl96}.  In computing these models, we assumed the stellar emission comes from a centrally located point source.  Considering the simple geometry of the compact starburst evident in the WFPC2 UV image, namely a single, compact knot \citep{bart98}, this represents a good approximation to the actual morphology.

     Our method in this analysis was to find the simplest model that best reproduces the emission line spectrum observed in NGC~4569.  Thus, we sought the fewest number of emission line gas components that combine to match the spectrum.  Each model gas ``component" was specified by a constant hydrogen gas density, n$_H$, and distance from the ionizing starburst, R.  Combined with the stellar luminosity, these parameters determine the ionization parameter for each component, 
\begin{displaymath}
U=4\pi L / n_{H}R^{2}c,
\end{displaymath}
which determines the ionization state of the gas.

     Modeling the emission line region as one or a few discrete components is a simplification.  The distribution of emission line clouds is certainly more complex than forming a set of distinct, perfect shells centered on the starburst.  However, the minimal component method is the most instructive, since it reveals the general physical conditions that are required to generate the observed emission line spectrum.

     All models were computed for an open geometry, {\it i.e.}, all emission escapes the clouds without interacting with emission line gas on the opposite side of the starburst.  This assumes {\it a~priori} a covering factor less than unity.  Furthermore, all emission line clouds were assumed to be matter bounded, with an arbitrarily large column of neutral gas in the backside of the clouds.

     The WFPC2 [\ion{O}{3}] and H$\alpha$ flux measurements by \citet{pogg00} provide constraints on the distribution of the emission line gas.  They found that $\sim$30\% of H$\alpha$ and $\sim$75\% of the [\ion{O}{3}] flux is emitted in the inner R=20~pc region of the nucleus (see \S~2.1).  The electron density of the [\ion{S}{2}] emitting region can be determined from the ratio of the [\ion{S}{2}]~$\lambda$6716 to [\ion{S}{2}]~$\lambda$6731 line fluxes \citep{oste89}.  In NGC~4569, F$_{[S II]~\lambda 6716}$/F$_{[S II]~\lambda 6731}$=1.0$\pm$0.2, which indicates an electron density of n$_e$=600$^{+300}$$_{-200}$~cm$^{-3}$.  This is well below the critical density of these lines, n$_c$$\sim$10$^4$~cm$^{-3}$, and therefore provides an important diagnostic for photoionization analysis.

\subsection{Model Results}

     We computed a series of models with the 5~Myr starburst SED (Figure~5) at 
R=20~pc, consistent with the size of the [\ion{O}{3}]~$\lambda$5007 emission 
core derived from the WFPC2 imagery.   Guided by our population 
synthesis results, we assumed Z=Z$_\sun$/2.5 for the nebular gas, where 
Z$_\sun$ is from \citep{grev89}, modified by depletions onto dust grains.  Gas-phase depletions and dust grain abundances typical of the Galactic 
ISM \citep[and references therein]{ferl96}, scaled by a factor of 1/2.5 
to correct for sub-solar metallicity, were assumed.  The resulting 
abundances relative to hydrogen are:\\
He=1.00x10$^{-1}$, C=5.68x10$^{-5}$, N=3.73x10$^{-5}$, O=1.78x10$^{-4}$, Ne=4.68x10$^{-5}$, Mg=3.04x10$^{-6}$, Al=1.18x10$^{-8}$, 
Si=4.28x10$^{-7}$, S=6.48x10$^{-6}$, Ca=9.16x10$^{-8}$, and 
Fe=1.30x10$^{-7}$.\\
Results for n$_H$=10$^3-$10$^6$~cm$^{-3}$ are given in models A1$-$A4 in 
Table~3. 
     
     The observed [\ion{O}{3}]~$\lambda$5007/H$\beta$ flux ratio in the inner 20 pc core of the nucleus is 
matched with n$_H$$\sim$10$^4-$10$^5$~cm$^{-3}$.  The [\ion{O}{1}]~$\lambda$6300 luminosity can 
be generated if a high density component, n$_H\sim$10$^6$~cm$^{-3}$, is 
also present (model~A4).  However, the other LINER features are not 
matched by these models due to the high ionization parameter.  For example, 
the predicted [\ion{S}{2}]~$\lambda$6726/H$\beta$ ratio in model~B1 is only 
10\% of the observed value and [\ion{N}{2}]~$\lambda$6584/H$\beta$ is 
underestimated by a factor at least two in all models.  We find that no 
choice of n$_H$ or gas-phase abundances can reproduce these lines in the inner 20~pc region.

     Thus, if our derived luminosity and SED for the starburst is 
correct, much of the [\ion{S}{2}]~$\lambda$6726 and [\ion{N}{2}]~$\lambda$6584 
emission must arise at further distances from the central starburst.  In model~B of Table~3, results are given for R=80, 130, and 180~pc.  To 
satisfy the constraint imposed on the electron density by the 
[\ion{S}{2}]~$\lambda$6716/[\ion{S}{2}]~$\lambda$6731 ratio, we assumed 
n$_H$=10$^3$~cm$^{-3}$.  These results indicate R$>$130~pc is required to generate the [\ion{S}{2}]~$\lambda$6726 and [\ion{N}{2}]~$\lambda$6584 flux without 
overestimating the [\ion{O}{3}]~$\lambda$5007 emission.  Only at these large 
distances is the ionizing photon flux sufficiently dilute.  However, 
this gas would lie near or beyond the edge of the area sampled by the 
aperture.  Thus, any model of this extended emission region must account for aperture coverage effects.
 
     We have also computed photoionization models using the 6~Myr SED.  
Results are given in models C and D, which were computed with the same 
parameters as models A and B, respectively.  Due to the decrease in photoionizing flux between the \ion{He}{2} and \ion{He}{1} edges compared to the 5~Myr SED (see Figure~5), the 6~Myr model SED generates much weaker [\ion{O}{3}] emission.  The [\ion{O}{1}]~$\lambda$6300 emission in the 6~Myr high density model (C4) is 
also much weaker because the SED is too steep to generate a sufficiently
deep partially ionized zone in the backside of the cloud.  For the same reason, the [\ion{S}{2}] and [\ion{N}{2}] emission in the extended emission line region (model D) is weaker than in the 5~Myr model. 
 
{\it Best-Fit Model for the Emission Line Spectrum}

     In Table~4, we present a composite model that represents our 
best-fit to the emission line spectrum of NGC~4569 using the 5~Myr SED.  A minimum of three distinct components is required for an acceptable fit.  For component~1, we set n$_H$=10$^{4.5}$~cm$^{-3}$ at R=20~pc to reproduce the [\ion{O}{3}]~$\lambda$5007/H$\beta$ ratio measured in the core from the WFPC2 imagery.  Component~2 consists of a low density gas at a relatively large distance from the ionizing starburst (low U), which is required to
produce the [\ion{S}{2}] emission and much of the [\ion{N}{2}] flux.  We find that n$_H$$\leq$900~cm$^{-3}$ is needed to reproduce the [\ion{S}{2}]~$\lambda$6716/[\ion{S}{2}]~$\lambda$6731 flux ratio in the composite model.  A high density component (component~3) is required to generate the [\ion{O}{1}]~$\lambda$6300 luminosity.
We chose component~3 to be at the location of the [\ion{S}{2}] emitting gas, however, this component could be located closer to the starburst, as illustrated in model~A4 in Table~3.  The density for this 
component, n$_H$=10$^{5}$~cm$^{-3}$, was chosen to maximize 
[\ion{O}{1}]~$\lambda$6300 and [\ion{N}{2}]~$\lambda$6584 to provide the best overall fit.  This composite model also satisfies the upper limits measured for the UV emission lines given in Table~1.  Specifically, the predicted \ion{He}{2}~$\lambda$1640 flux, listed in Table~4, is equals the upper limit.

     In deriving this model, the relative contribution of each component to the total emission was chosen to best match the observed spectrum.  The covering factors for the three gas components were then 
derived by setting the sum of their H$\beta$ luminosities equal to the 
observed value, L$_{H\beta}=$8x10$^{39}$~ergs~s$^{-1}$.  We corrected 
for the aperture size in determining the total covering factors of the 
two extended components.  Assuming that the emission from these 
components comes from a shell of emission at R=180~pc from the 
starburst, we find that the 160~pc~x~320~pc aperture samples a total of 
50\% of this emission.  Thus, the covering factors listed in Table~4 
equal the H$\beta$ luminosity derived for each component, divided by the H$\beta$ luminosity predicted by the models, divided by the fraction of 
the total emission sampled by the aperture.  This is only a rough correction for aperture coverage, since it assumes that a uniform gas surrounds the starburst and that the aperture is centered on this emitting shell.

     The high covering factor derived for our model indicates we may be observing the central starburst through the emission line gas.  This may be responsible for the large gas column density in our line-of-sight to the starburst revealed by the strong interstellar absorption features and heavy extinction evident in the UV spectrum.  

     Although the above model is a simplification, the three components 
represent physically distinct zones of emission.  For example, the 
[\ion{S}{2}] and [\ion{O}{3}] components must be isolated and spatially distinct, with little line emitting gas between the two regions.  Otherwise, the [\ion{O}{3}]~$\lambda$5007/H$\beta$ and [\ion{S}{2}]~$\lambda$6726/H$\beta$ ratios cannot be simultaneously fit.  Furthermore, the large difference in 
densities required for the [\ion{O}{1}] and [\ion{S}{2}] regions indicates these lines arise in distinct clouds.

\section{Discussion}

\subsection{Implications for the Starburst SED}

     The results of our photoioinization modeling impose limits on the shape of 
the integrated SED of the central starburst in NGC~4569.  From models A and B in Table~3, we find that the actual SED cannot be much harder than the 5~Myr model.
Due to the high luminosity of the starburst and the low density of the 
\ion{S}{2} emitting gas, a harder SED cannot reproduce the \ion{S}{2}/\ion{O}{3}
flux ratio and simultaneously satisfy the constraint on the location of the \ion{O}{3} gas derived from the WFPC2 imagery - the ionization parameter is too high given the small aperture used to obtain the spectrum.  

     This implies that models with very large populations of W-R stars, such as the solar metallicity 5 and 6~Myr models generated with STARBURST99, are not correct for NGC~4569 since they have very hard SEDs.  This is consistent with our limit derived on the W-R population and metallicity in \S~3.2.  Similarly, stellar atmosphere models that predict stronger flux beyond the \ion{He}{1} ionization edge for O and B stars than the \citet{leje97} models will produce a worse fit to NGC~4569.

     Conversely, the 6~Myr SED, with only a weak contribution from hot W-R stars, is too steep to generate the LINER features (see models C and D in Table~3).  It lacks the extended tail of hard photons that is required to generate a warm partially ionized zone in the backside of the clouds where these lines are emitted strongly.  The 6~Myr model represents a typical \ion{H}{1} region SED.  This indicates that W-R stars must contribute to some extent to the ionization in NGC~4569, as claimed by \citet{bart00}.

\subsection{Implications of Uncertainties on the Derived Model}

    The uncertainty in the distance, D, to NGC~4569 introduces an uncertainty in the absolute values of the measured and derived parameters ({\it e.g.}, luminosities and spatial extent of the emission line gas, R).
However, since R~$\propto$~D and L~$\propto$~D$^2$, the ionization parameter is independent of D (U~$\propto$~L/R$^2$).  For a given ionization parameter, photoionization calculations show that the emission line luminosities scale proportionally with the ionizing luminosity, except for models with very high U, which are not applicable to NGC~4569.  Thus, a change in D does not affect the essential characteristics of our photoionization modeling results, namely, the derived densities, covering factors, and geometry of the emission line components.

    The largest uncertainty in our modeling results is due to the uncertainty in the extinction law, and thus ionizing luminosity, for the starburst in NGC~4569.  We have tested various extinction curves from the literature.  There are several observational constraints that place limits on the steepness of the extinction curve in NGC~4569.  Specifically, the unreddened starburst cannot exceed the optical continuum luminosity we derived for the inner core region using WFPC2 imagery (\S~2.3) and the upper limit on the bolometric luminosity implied by the far-infrared data (\S~2.4).  These luminosity constraints provide a lower bound on the steepness of the extinction curve.
The \citet{sava79} curve, which we employed for our analysis, satisfies both of these constraints.
However, we find that the \citet{calz94} extinction law, which was empirically derived for starburst galaxies, greatly overestimates these luminosities and is thus too ``grey" to be applied to NGC~4569. 
Conversely, curves that are much steeper than the \citet{sava79} curve underestimate the ionizing luminosity required to generate the emission line luminosity, as evidenced by the high covering factor derived from our photoionization models. 
For example, the SMC curve of \citet{hutc82} results in a factor of two deficit in ionizing photons, and the very steep curve derived for the Seyfert~1 galaxy NGC~3227 by \citet{cren01} gives an even worse fit.  Additionally, these steep curves would require that the emission line extinction, as determined from the Balmer decrement, is much larger than the extinction of the starburst.

\subsection{Physical Interpretation of the Model Results}

      As demonstrated in \S~4.2, a specific geometry for the nuclear gas is required if we are to accept stellar photoionization as the sole ionization and excitation mechanism in NGC~4569.  A complete model, therefore, must provide a physical interpretation for the distinct spatial components.  

     The mechanical energy due to frequent supernovae and strong stellar winds associated with the young, massive stellar population may create this structure.  In this scenario, the [\ion{S}{2}] emitting gas in NGC~4569, component~2 in our best-fit model, may be swept-up interstellar gas surrounding the evacuated hot cavity of an interstellar supershell.  The more compact [\ion{O}{3}]~$\lambda$5007 clouds revealed in the WFPC2 narrowband imagery may be remnants of the original molecular cloud from which the starburst was born.  In agreement with this, star-forming molecular clouds are thought to consist of dense condensations ({\it e.g.}, n$_H\sim$10$^2$-10$^5$~cm$^{-3}$) embedded in very tenuous intervening gas \citep[and references therein]{efst00}.  The diffuse component would be swept up relatively rapidly by the mechanical energy input by the young stars and may lead to the bi-component spatial structure proposed here.

     To assess the validity of this model, we calculated the initial ambient density that would be required to produce the line luminosity of the putative extended [\ion{S}{2}] emitting region, assuming this gas is comprised of swept-up, diffuse material.  Using L$_{H\beta}$=4x10$^{39}$~erg~s$^{-1}$, which was derived for component~2 in our best-fit model, a minimum of 2x10$^{61}$ hydrogen atoms are necessary to produce the emission.  This is a lower limit to the total gas since it accounts only for the line emitting gas.  Dividing this total by the volume enclosed in an R=180~pc region (consistent with our best-fit model), an average hydrogen density of n$_H$=0.02~cm$^{-3}$ is found for the tenuous component of the pre-starburst interstellar gas.  This is typical of a diffuse interstellar density.  

     The constraints derived on the location of the [\ion{S}{2}] gas from our modeling results are consistent with the extent of the nuclear radio emission (see discussion in \S~2.4).  Thus, the extended radio flux could be free-free emission from the [\ion{S}{2}] emitting nebular gas or non-thermal emission generated in shocks at the interface of the swept-up interstellar medium.  Additionally, if we assume a typical temperature and density for the hot, tenuous cavity implied by this model, {\it e.g.}, T=10$^6$~K and n$_H\leq$10$^{-2}$~cm$^{-3}$, we find that the X-ray luminosity predicted by the hot plasma models of \citet{raym76} is much weaker than the {\it ROSAT} measurement of the nucleus of NGC~4569.  Thus, the X-ray emission is likely dominated by X-ray binaries and/or supernova remnants.

     Therefore, all observations are consistent with this model.  We conclude that it provides a viable explanation for the bi-component gas required by our photoionization modeling, and that the stellar photoionization hypothesis remains a plausible solution for NGC~4569.

\section{Summary}

     Our spectral population synthesis modeling results indicate the nuclear starburst in NGC~4569 is 5-6~Myr old and implies a relatively brief duration of star formation.  The lack of detectable \ion{He}{2}~$\lambda$1640 emission constrains the population of W-R stars and implies a sub-solar metallicity, with $Z\la$Z$_\sun$/2.5.  Luminosity constraints limit the steepness of the extinction curve that applies to NGC~4569.  The Galactic curve of \citet{sava79} satisfies all luminosity constraints, and gives a similar reddening for the starburst and nebular line emission, $E(\bv)=$0.50.  These results imply extreme conditions in the nuclear starburst, with $\sim$5x10$^4$ O and B stars compacted in the inner 9\arcsec~x~13\arcsec region of the nucleus.
Additionally, we find that this young, very dense starburst is distinct from the older, more spatially extended population of A supergiants detected in the optical spectrum \citep{keel96}.  

     The results of our photoionization modeling build on the previous analysis of \citet{bart00}, which demonstrated the general conditions required for a starburst to produce the enhanced LINER features observed in NGC~4569.  Using all available observational constraints on the density, luminosity, and geometery of the nuclear region in our analysis, we have identified the physical elements that are required if the transition object spectrum in NGC~4569 is generated solely by stellar photoionization:\\
(a) At least two spatially distinct line emitting components are needed - a very compact region emitting [\ion{O}{3}]~$\lambda$5007 and a more extended component for [\ion{S}{2}]~$\lambda\lambda$6716,6731.  We propose a model in which these components correspond to the remnants of the parent molecular cloud of the starburst and to a swept-up ``shell" of interstellar material, respectively.  \\
(b) A high density (n$_H$$\sim$10$^6$~cm$^{-3}$) gas is required to explain the [\ion{O}{1}]~$\lambda$6300 emission.\\
(c) The 5~Myr, Z=Z$\sun$/2.5 model SED derived with STARBURST99 \citep{leit99} must be similar to the actual flux distribution of the starburst.  If our derived extinction is correct, luminosity constraints limit the SED from being much harder.  Conversely, an SED that is much softer cannot generate the LINER features.  These results are consistent with our population synthesis results, which indicate Z$_\sun$/10$<Z\la$Z$_\sun$/2.5.  Thus, our photoionization models support the assertion by \citet{bart00} that W-R stars play an important role in the ionization of NGC~4569.  

    In conclusion, we have shown through detailed modeling that photoionization by the nuclear starburst is capable of generating the observed LINER/\ion{H}{2} region transition object spectrum, consistent with all observations.  Our model places specific requirements on the physical conditions in the nucleus.  Thus, it makes several predictions that can be tested with future observations.

    Facilities of the Laboratory for Astronomy and Solar Physics at NASA/GSFC were used in performing this research.  
We thank the referee, W. Schmutz, for providing very helpful comments for this paper.  We also thank Steve Kraemer and Mike Crenshaw for useful discussions and advice, Margaret Smith Neubig for assisting with the population synthesis analysis, and Gary Ferland (CLOUDY) and Claus Leitherer (STARBURST99) for making their software publicly available.

\clearpage

\figcaption{{\it HST}/WFPC2 continuum image of NGC~4569 using the F547M filter.  The square delineates a 0$\arcsec$.35~x~0$\arcsec$.35 region corresponding approximately to the spatial extent of the compact starburst.  For the adopted distance to NGC~4569, each side of the square represents about 25~pc.\label{fig1}}

\figcaption{Multi-wavelength spectrum of the nucleus of NGC~4569.  The 6~cm and 2~cm VLA radio data ($\ast$) are upper limits placed on any unresolved (1$\arcsec$) core emission from \citet{neff92}.  The {\it IRAS} data points ($\diamond$) are from \citeauthor{neff92} and sample the inner $\approx$1$\arcmin$. The UV (dot-dashed line) is from the {\it HST}/FOS spectrum.  \citet{bart98} found the UV source has a FWHM size of $\approx$0$\arcsec$.15.  Optical data (solid line) are for the inner 0$\arcsec$.35 core of emission measured from the F547M WFPC2 image (this study).  The X-Ray data (dashed line) is a power law fit to the observed {\it ROSAT}/HRI data, which has $\approx$4$\arcsec$ resolution, and thus represents an upper limit to the inner core luminosity.  Upper limits are indicated with arrows. \label{fig2}}

\figcaption{Population synthesis spectra for various ages plotted for the (a) \ion{N}{5}~$\lambda$1240, (b) \ion{Si}{4}~$\lambda$1400, (c) \ion{C}{4}~$\lambda$1550 and (d) \ion{He}{2}~$\lambda$1640 spectral regions.  Both the theoretical (solid line) and observed spectrum (dashed line) have been normalized.  Empirical fits to the broad wind absorption features are plotted on the spectra for \ion{Si}{4} and \ion{C}{4} (smooth line).  All model spectra were generated with CLUST\_FLUX using $\xi_{MS}$ and Z$_\sun$.  The prominent stellar wind and photospheric features are identified above the spectrum.  Strong interstellar lines are marked (X) below the spectrum.  The spectrum of NGC~4569 has been convolved (smoothed) with a $\sigma$=1~\AA\  Gaussian filter.  All model spectra were smoothed with a $\sigma$=1.5~\AA\  filter to best match the resolution of the NGC~4569 data presented in the figure. \label{fig3}}

\figcaption{Population synthesis spectra for 1890-1990~\AA\.  The model spectra (solid line) and the observed spectrum (dashed line) have been normalized.  The \ion{Fe}{3} lines are identified at the top of the 0 and 5~Myr models with tick marks.  The spectra were smoothed as described in the caption of Figure 3.  \label{fig4}}

\figcaption{5 Myr (solid line) and 6 Myr (dashed line) model starburst spectral energy distributions.  The SEDs were derived with Z=Z$_\sun$/2.5 and the Miller \& Scalo (1979) IMF, assuming an instantaneous burst of star-formation using the population synthesis code STARBURST99 (Leitherer et al. 1999).\label{fig5}}

\clearpage

\begin{deluxetable}{lrr}
\tablecaption{Emission Line Data \label{tbl-1}}
\tablewidth{0pt}
\tablehead{
\colhead{Line} & \colhead{Observed\tablenotemark{a}}   
& \colhead{Dereddened\tablenotemark{b}}}
\startdata
H$\beta \lambda$4861 & 1.0\tablenotemark{c} & 1.0\tablenotemark{d} \\
$[$\ion{O}{3}] $\lambda$5007 & 1.2 ($\pm$ 0.2) &  1.1 ($\pm$ 0.5) \\
$[$\ion{O}{1}] $\lambda$6300 & 0.31 ($\pm$ 0.10)  & 0.19 ($\pm$ 0.08) \\
H$\alpha$ $\lambda$6563 & 5.0 ($\pm$ 1.3) & 2.9 ($\pm$ 1.1) \\
$[$\ion{N}{2}] $\lambda$6583 & 4.5 ($\pm$ 1.0) & 2.6 ($\pm$ 0.9) \\
$[$\ion{S}{2}] $\lambda\lambda$6716,6726 & 2.0 ($\pm$ 0.4) & 1.2 ($\pm$ 0.4)\\ 
\ion{C}{4} $\lambda\lambda$1548,1551 & $\leq$0.04 & $\leq$0.3 \\
\ion{He}{2} $\lambda$1640 & $\leq$0.04  & $\leq$0.3 \\
\ion{C}{3}] $\lambda$1909 & $\leq$0.02 & $\leq$0.2 \\
\ion{C}{2}] $\lambda$2326 & $\leq$0.02 & $\leq$0.2 \\
$[$\ion{O}{2}] $\lambda$2470 & $\leq$0.02 & $\leq$0.1 \\
\enddata
\tablenotetext{a}{Optical fluxes from Ho et al. (1997a); UV upper limits from Maoz et al. (1998); Fluxes relative to H$\beta$}
\tablenotetext{b}{Corrected for E(B-V)=0.48, Savage \& Mathis (1979) extinction curve}
\tablenotetext{c}{Observed: L$_{H\beta}$=1.6x10$^{39}$ erg s$^{-1}$; assuming D = 16.8 Mpc}
\tablenotetext{d}{Dereddened: L$_{H\beta}$=8.1x10$^{39}$ erg s$^{-1}$; assuming D = 16.8 Mpc}
\end{deluxetable}
\clearpage

\begin{deluxetable}{lrrrr}
\tablewidth{0pt}
\tablecaption{FOS Observations of NGC~4569 \label{tbl-2}}
\tablehead{
\colhead{Data Set} & \colhead{Grating} & \colhead{Aperture} &
\colhead{Resolution} & \colhead{Exposure}\\
\colhead{} & \colhead{} & \colhead{$\arcsec$} & 
\colhead{\AA } & \colhead{sec}}
\startdata
Y3400104T  &  G130H  & 0.86 &  $\sim$1.0 & 2370 \\
Y3400105T  &  G130H  & 0.86 &  $\sim$1.0 & 2420 \\
Y3400106T  &  G190H  & 0.86 &  $\sim$1.5 & 1320  \\
Y3400107T  &  G270H  & 0.86 &  $\sim$2.0 & 690 \\
\enddata
\end{deluxetable}
\clearpage

\begin{deluxetable}{lrrrrrrrrrr}
\tabletypesize{\scriptsize}
\tablewidth{0pt}
\tablecaption{Photoionization Model Results \label{tbl-3}}
\tablehead{
\colhead{Model} & \colhead{SB Age\tablenotemark{a}} & \colhead{R} & \colhead{n$_H$} & \colhead{Log(U)} & \colhead{L$_{H\beta}$} & \colhead{[O III] $\lambda$5007 \tablenotemark{b}} & \colhead{[O I] $\lambda$6300} & \colhead{[N II] $\lambda$6583} & \colhead{[S II] $\lambda$6726 \tablenotemark{c}} & \colhead{R$_{[S II]}$\tablenotemark{d}}\\
\colhead{} & \colhead{Myr} & \colhead{pc} & \colhead{cm$^{-3}$} & \colhead{} &
\colhead{10$^{40}$ erg s$^{-1}$} & \colhead{} & \colhead{} & 
\colhead{} & \colhead{} & \colhead{}}
\startdata
A1 & 5 & 20 & 10$^3$ & -1.5 & 0.97 & 6.6 & - & 0.19 & 0.11 & 0.84 \\
A2 &   & &  10$^4$ & -2.5 & 1.7 & 4.6 & 0.01 & 0.95 & 0.25 & 0.70 \\
A3 &  &  &  10$^5$ & -3.5 & 2.0 & 0.78 & 0.12 & 1.2 & 0.24 & 0.50 \\
A4 &  &  &  10$^6$ & -4.5 & 2.1 & - & 0.47 & 0.38 & 0.10 & 0.40 \\
B1 & 5 & 80 & 10$^3$ & -2.7 & 1.0 & 3.5 & 0.02 & 1.3 & 0.77 & 0.84 \\
B2 &   & 130 & 10$^3$ & -3.1 & 2.1 & 2.0 & 0.04 & 1.8 & 1.3 & 0.84 \\
B3 &  & 180 & 10$^3$ & -3.4 & 2.1 & 0.83 & 0.09 & 2.1 & 1.8 & 0.84 \\
C1 & 6  & 20 & 10$^3$ & -1.4 & 1.8 & 2.0 & - & 0.79 & 0.10 & 0.84 \\
C2  & &  &  10$^4$ & -2.4 & 2.1 & 0.77 & 0.01 & 1.4 & 0.24 & 0.70 \\
C3  & &  &  10$^5$ & -3.4 & 2.5 & 0.07 & 0.04 & 0.93 & 0.13 & 0.50 \\
C4  & &  &  10$^6$ & -4.4 & 2.6 & - & 0.16 & 0.18 & 0.03 & 0.40 \\
D1 & 6 & 80 & 10$^3$ & -2.6 & 2.3 & 0.43 & 0.01 & 1.4 & 0.65 & 0.84 \\
D2  & & 130 & 10$^3$ & -3.1 & 2.4 & 0.15 & 0.02 & 1.5 & 1.0 & 0.84 \\
D3  & & 180 & 10$^3$ & -3.3 & 2.4 & 0.07 & 0.03 & 1.5 & 1.3 & 0.84 \\
\enddata
\tablenotetext{a}{Model starburst SED from Leither et al. (1999)}
\tablenotetext{b}{All fluxes relative to H$\beta$}
\tablenotetext{c}{Total [S II] $\lambda$6716+$\lambda$6731 flux}
\tablenotetext{d}{R$_{[S II]}$=L$_{[S II] \lambda6716}$/L$_{[S II] \lambda6731}$}
\end{deluxetable}
\clearpage

\begin{deluxetable}{lrrrrrrrrrr}
\tabletypesize{\scriptsize}
\tablewidth{0pt}
\tablecaption{Best-Fit Model using the 5 Myr Starburst SED \label{tbl-4}}
\tablehead{
\colhead{} & \colhead{R}  & \colhead{n$_H$} & \colhead{Log(U)} &
\colhead{[O III] $\lambda$5007\tablenotemark{a}} & \colhead{[O I] $\lambda$6300} &
\colhead{[N II] $\lambda$6583} & \colhead{[S II] $\lambda$6726 \tablenotemark{b}} & \colhead{He II $\lambda$1640} & \colhead{R$_{[S II]}$\tablenotemark{c}} & \colhead{Covering Factor}\\
\colhead{} & \colhead{pc}  & \colhead{cm$^{-3}$} & \colhead{} &
\colhead{} & \colhead{} & \colhead{} & \colhead{} & \colhead{} & \colhead{}}
\startdata
Comp 1 & 20  & 10$^{4.5}$ & -3.0 & 2.9 & 0.04 & 1.3 & 0.26 & 0.41 & 0.5  & 0.14\\
Comp 2 & 180 & 900 & -3.4 & 0.83 & 0.09 & 2.1 & 1.8 & 0.30 & 0.84 & 0.38\\
Comp 3 & 180 & 10$^5$ & -5.4 & - & 1.1 & 1.1 & 1.6 & 0.01 & 0.54 & 0.13\\
Composite\tablenotemark{d} &  &  &  & 1.3 & 0.24 & 1.8 & 1.3 & 0.30 & 0.80 &  \\
Observed\tablenotemark{e} &  &  &  & 1.1($\pm$0.5) & 0.19($\pm$0.08) & 2.6($\pm$0.9) & 1.2($\pm$0.4) & $\leq$0.3 & 1.0($\pm$0.2) & \\
\enddata
\tablenotetext{a}{All fluxes relative to H$\beta$}
\tablenotetext{b}{Total [S II] $\lambda$6716+$\lambda$6731 flux}
\tablenotetext{c}{R$_{[S II]}$=L$_{[S II]\lambda6716}$/L$_{[S II] \lambda6731}$}
\tablenotetext{d}{L$_{H\beta}$ comprised of 33\% Comp 1, 50\% Comp 2, 17\% Comp 3}
\tablenotetext{e}{Dereddened by E(B-V)=0.48 using Savage \& Mathis curve}
\end{deluxetable}
\clearpage

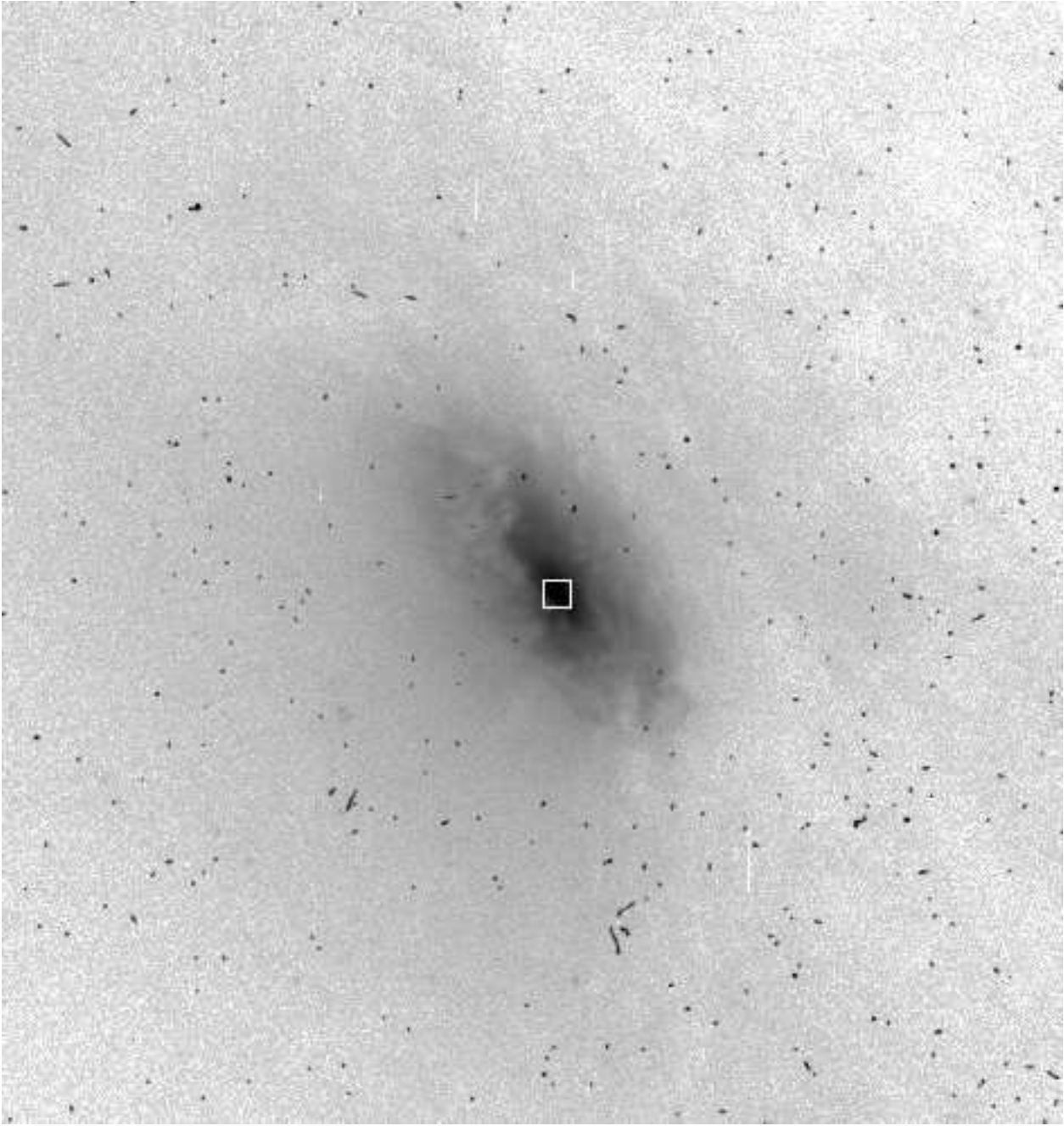
\end{document}